\documentclass{kapproc} 
 
\usepackage{procps}  
 
\usepackage[dvips]{graphicx} 
 
\upperandlowercase 
 
\let\footnote\savefootnote 
 
\kluwerbib 
 
\usepackage{epsfig}

\begin{document} 
 
\articletitle{Close by Compact Objects and Recent Supernovae in the Solar
Vicinity} 
 
 
 
\author{Sergei Popov\altaffilmark{1,2} 
} 
 
\altaffiltext{1}{University of Padova\\ 
via Marzolo 8, 35131,
Padova, Italy} 
 
\altaffiltext{2}{Sternberg Astronomical Institute\\ 
Universitetski pr. 13, 119992, Moscow, Russia} 
\email{polar@sai.msu.ru} 
 
\begin{abstract} 
 I discuss young close-by compact objects, recent supernovae
in the solar neighbourhood, and point to their connection
with cosmic ray studies. Especially the role of the Gould Belt is
underlined.
\end{abstract} 
 
 
\section*{Introduction}

In this short paper at first I shall try to show links between studies 
of cosmic rays (CRs) and studies of close-by compact objects (in particular
neutron stars -- NSs). 
The reason for an existence of such relation is
obvious: both phenomena (CRs and NSs) have the same origin -- 
supernova (SN) explosions. Then I discuss in more details
the population synthesis of close-by NSs. 
The analysis of the population of these
sources makes us to conclude that the solar neighbourhood 
(by that I mean a region about few hundred of parsecs around the Sun)
in enriched with  young NSs. 
It is a natural consequence of the existence of the Gould Belt --
local structure formed by massive stars. 

 The main message I want to deliver in this contribution is the following.
We are living in a region of the Milky way enriched with massive stars
in comparison with a typical place at 8 kpc from the center of the Galaxy.  
Due to this fact SN rate in $\sim$~600 pc around the Sun during the last
few tens of Myrs is enhanced. It results also in the 
ebhanced number of near-by young compact objects. Now we know about two tens 
of young (age $<$ 4 Myrs) close (distance $<$ 1 kpc) NSs. 
This local (in space and time) increase of SN rate 
can be important for modeling of CR flux on the Earth.

\section{Cosmic rays and recent close-by supernovae} 

 Galactic CRs are genetically connected with SN:
CRs are accelerated in SN remnants (see other materials in this volume, 
especially contributions 
by  Michal Ostrowski,  Vladimir Ptuskin and  John Wefel).
The local CR density is inevitably dependent on the rate of SN in the Galaxy
in general and in the solar proximity in particular.
CR flux can be increased if the Sun 
appears in  a region of the Galaxy with enhanced SN rate, or if by chance
several SN explosions in a row happen in the vicinity of the Sun. 

For example Shaviv (2004) suggested, that passages of the Sun through
galactic spiral arms are accompanied by increase of CR flux on Earth
(at this conference some results were presented in the talk by
Smadar Levi). As CRs can have influence on Earth climate (in particular on
cloud formation)
such passages can be responsible for the (quasi)periodic appearence of ice age
epochs on our planet.  However, I would like to note, that even local
quasiperiodic fluctuations of starformation rate (and correspondent changes
in SN rate) can give a similar effect. 
Such fluctuations were found by
Vanbeveren and De Donder (2003) in their modeling of star formation
in the solar vicinity. 
Another possibility to enhance SN rate is connected with the 
formation of structures like the Gould Belt 
(see description of the Belt below). 
\footnote{Effects of the Belt on CR characteristics were 
discussed in Pohl et al. (2003) and in Dermer (2004). } 
All these effects can work together, and therefore
modeling of the CR flux history 
becomes a very complicated business because of a ``turbulent history'' of 
star formation in $\sim$ 1 kpc around the Sun. 

For a long time local SN were considered as one of candidates for 
a catastrophic
event responsible for  mass extinction of species. Still SN potentially can 
play a ``positive'' role in the history of life on Earth.
An interesting possibility of SN influence on bioworld of our planet
can be connected
with the hypothesis discussed by Tsarev (1999; it is also described in
an easier available paper by Chernavskii 2000). In the original version the
author suggested that chirality symmetry breaking 
existing in the bioorganic world can be  connected with strong neutrino
burst due to a close SN which happened in the early stages of the history
of the Earth. 
However, a close-by SN (close enough to povide a strong neutrino flux)
can just destroy terrestrial bioorganics. 
In that sense  it is possible to suggest
a more promising hypothesis which  includes so
called "choked gamma-ray bursts" (Meszaros, Waxman 2001). 
If a jet is unable to penetrate through the stellar envelope, then only 
a neutrino beam will break-out.  In that case it is possible both: 
to increase the neutrino flux, 
and to decrease negative influence of hard radiation of a SN on the biosphere. 

 Recent local SN (and related close-by NS)
can strongly influence not only the flux of  CR but  the spectrum as well.
Erlykin and Wolfendale (1997) proposed that a knee in the spectrum
is due to a single source. 
According to their estimates there should exist
a near-by relatively young SN remnant.
An estimated age of such a hypothetical object was found
by these authors to be equal to 80-100 kyr, and its distance -- to 230-350 pc.
They suggested that the Monogem nebula 
(and related to it PSR 0656+14) can be such
a source which produce the knee in the spectrum (see also the contribution
by Jorg Hoerandel in this volume for a more detailed discussion of the knee).  

How probable an appearence of such an object is?
If the Sun is sutuated in a typical region of the Galaxy at 8 kpc from its 
center then the probability of appearence of a SN remnant of necessary age
at necessary distance is very low. However, we are living in a region
of enhanced SN rate, and there are many young compact objects around us.
In the following section I am going to discuss this population
and it origin in more details. 

\section{Young close-by NSs and the Gould Belt} 

 The solar proximity is rich in young NSs. 
Some of them represent a new class of sources:
radioquiet X-ray dim isolated NSs (see a review in Treves et al. 2000).
The reason of such enrichment is an existence of a structure formed by
massive stars -- the Gould Belt. In the following subsection we 
discuss it in some details. Then we present new results on 
a population synthesis
of young NSs, and finally we discuss a convenient way to represent
this population.

 Population of near-by young NSs is presented in the table (reproduced
from Popov, Turolla 2004).
We included there 20 objects. Thirteen of them are detected as thermal 
emitters, i.e. as {\it coolers}. They are seven radioquiet ROSAT NSs 
(aka the Magnificent seven -- M7),
Geminga and geminga-like source 3EG J1835+5918, and four normal
radiopulsars (among which the famous Vela pulsar, and PSR B0656+14
which is probably connected with the single source, see above).
Seven close-by NSs which are not detected as thermal emitters 
(shown at the bottom of the table) are
normal radiopulsars with ages 2-4 Myrs.

\subsection{The Gould Belt}

The Gould Belt  is a structure
consisting of clusters of massive stars. About 2/3 of massive stars
inside 600 pc around the Sun belong to the Belt 
(see a recent paper by Grenier 2004 for main references).
The Gould Belt radius is about 300-500 pc. 
It is inclined at $\sim$18-22$^\circ$ respect to the galactic plane.
The Sun is situated not far from the center of  that disc-like structure. 
The age of the Belt is not well estimated, 30-60 Myrs can be a good guess.
The origin of this system is unclear. One of the possibility is an impact of a
high velocity cloud with the galactic disk. 
It is important to note, that 
the Sun and the Belt are not genetically related to each other.

Due to the presence of the Belt the rate of SN
around us (say in a few hundred parsecs) during last several tens of million
years is higher than it is at an average place at a solar distance from the
galactic center. The enhancement is about 3-4 times over the Galactic average.
Grenier (2003) estimates a SN rate in the Belt as 20-27 per Myr during the
last few Myrs.
Because of that there should be a local overabundance of
young NSs which can appear as hot cooling objects, as gamma-ray sources etc.
(in priciple in addition to NSs there should be young BHs in the solar 
neighbourhood, but these objects are much more elusive; it would be
very interesting to detect them, see Prokhorov, Popov 2002).
For example, such young NSs can form a subpopulation among EGRET unidentified 
sources. The existence of the Local Bubble can be also linked with the Belt
(Berghofer, Breitwerdt 2002).

\subsection{Log N - Log S distribution of close-by NSs}

One of the most powerful methods of investigating properties of some
population of sources is a Log N -- Log S distribution. It is an integral
distribution, which  shows the number of sources with a flux larger
than a given value. It is useful to remind two limiting cases.
If sources are distributited homogeniously and isotropicaly then the slope of
Log N -- Log S is equal to -3/2: the number of sources grows as the cube of a 
distance, but the observed flux decreases as the square.
In the case of a disk-like system the slope is -1 (the number of
objects is growing only as a square of a distance).
If sources are located at the same distance 
(for example, sources in a given far-away galaxy)
then Log N -- Log S shows distribution of sources in luminosity.

\begin{figure}[t] 
\epsfig{file=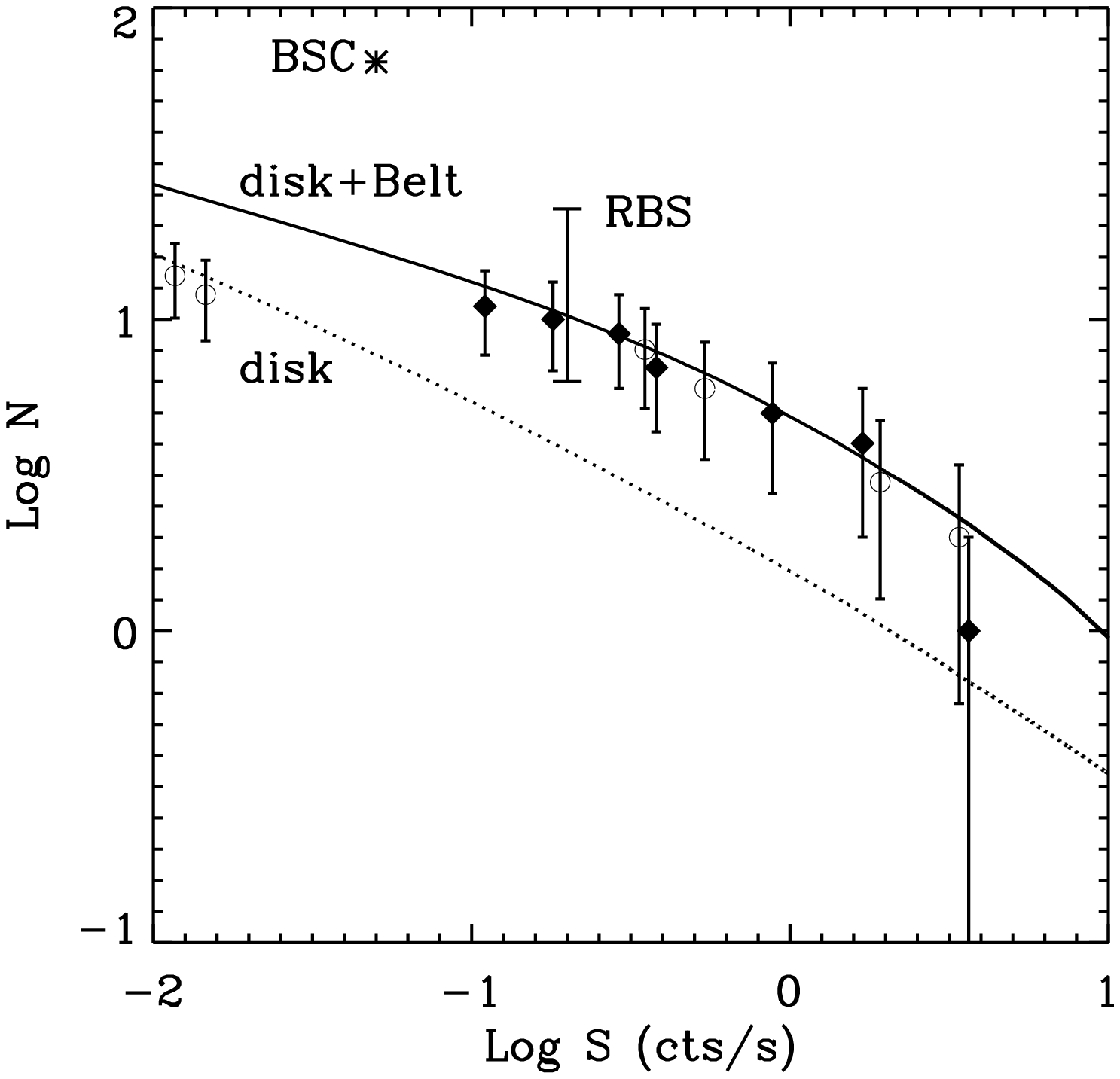, width=0.8\textwidth}
\caption{Log~N~--~Log~S distribution for close-by cooling isolated NSs. 
Black diamonds are plotted if the dimmest source at specified flux is
one of the "Magnificent seven". Otherwise we plot an empty circle.
Two lines represent results of calculation. Dotted line --- only stars
from the galactic disc contribute to the Log~N~--~Log~S distribution.
Solid line --- contribution of the Gould Belt is added. "RBS" and "BSC"
are two observational limits, obtained from the ROSAT data
(RBS: Schwope et al. 1999; BSC: Rutledge et al. 2003).
From Popov et al. (2003).}
\label{lognlogs}
\end{figure}

Popov et al. (2003) calculated Log~N~--~Log~S
distribution of cooling NSs in the solar vicinity, which can be observed
by ROSAT and other X-ray missions. The results are shown in the
fig.~\ref{lognlogs}.

\begin{table} 
\caption{Local ($D<1$ kpc) population of young (age $<4.25$ Myrs) 
isolated neutron stars\vspace*{1pt}} 
{\footnotesize 
\begin{tabular}{|l||c|c|c|c|c|c|} 
\hline 
\hline 
 
{} & {} & {} &{} &{} &{} &{}\\[-1.5ex] 
Source name & Period & CR$^a$ & $\dot P$ & D & Age$^b$ & Refs 
\\[1ex] 
            &   s     & cts/s & $10^{-15}$ s/s& kpc   & Myrs     & 
\\[1ex] 
 & & & & & & \\[1ex] 
\hline 
Magnificent seven & & & & & & \\[1ex] 
RX J1856.5-3754            &  ---  & 3.64  &  ---&0.117$^e$&$\sim0.5$& 
[1,2]\\[1ex] 
RX J0720.4-3125                 & 8.37 & 1.69  &$\sim 30-60$& ---&---& 
[1,3]\\[1ex] 
RX J1308.6+2127 & 10.3 & 0.29  & --- & --- & --- & 
[1.4] \\[1ex] 
RX J1605.3+3249       &  ---  & 0.88  & --- & --- & --- & 
[1]\\[1ex] 
RX J0806.4-4123                 &  11.37  & 0.38  & --- & --- & --- & 
[1,5]\\[1ex] 
RX J0420.0-5022                 &  3.45   & 0.11  & --- & ---   & --- & 
[1,11]\\[1ex] 
RX J2143.7+0654 &  ---    & 0.18  & --- & ---  & --- & 
[6]\\[1ex] 
 & & & & & & \\[1ex] 
\hline 
 Geminga type& & & & & & \\[1ex] 
PSR B0633+17            & 0.237 & 0.54$^d$ &10.97&0.16$^e$&0.34& 
[7]\\[1ex] 
3EG J1835+5918 & ---   & 0.015    & --- & ---  &  --- & 
[8]\\[1ex] 
 & & & & & & \\[1ex] 
\hline 
 Thermal. emit. PSRs & & & & & & \\[1ex] 
PSR B0833-45    & 0.089 & 3.4$^d$  & 124.88 & 0.294$^e$ & 
0.01& 
[7,9,10]\\[1ex] 
PSR B0656+14          & 0.385 & 1.92$^d$ &  55.01 & 0.762$^f$ & 0.11 
& 
[7,10]\\ [1ex] 
PSR B1055-52          & 0.197 & 0.35$^d$ &   5.83 & $\sim 1^c$ & 
0.54& 
[7,10]\\ [1ex] 
PSR B1929+10          & 0.227 & 0.012$^d$& 1.16 &  0.33$^e$  & 3.1& 
[7,10]\\[1ex] 
 & & & & & & \\[1ex] 
\hline 
Other PSRs & & & & & & \\[1ex] 
PSR J0056+4756  & 0.472 & --- & 3.57 &  0.998$^f$  & 2.1& 
[10]\\[1ex] 
PSR J0454+5543  & 0.341 & --- & 2.37 &  0.793$^f$  & 2.3& 
[10]\\[1ex] 
PSR J1918+1541  & 0.371 & --- & 2.54 &  0.684$^f$  & 2.3& 
[10]\\[1ex] 
PSR J2048-1616  & 1.962 & --- & 10.96&  0.639$^f$  & 2.8& 
[10]\\[1ex] 
PSR J1848-1952  & 4.308 & --- & 23.31&  0.956$^f$  & 2.9& 
[10]\\[1ex] 
PSR J0837+0610  & 1.274 & --- & 6.8  &  0.722$^f$  & 3.0& 
[10]\\[1ex] 
PSR J1908+0734  & 0.212 & --- & 0.82 &  0.584$^f$  & 4.1& 
[10]\\[1ex] 
\hline 
\hline 
\multicolumn{7}{l}{ 
$^a$) ROSAT PSPC count rate}\\ 
\multicolumn{7}{l}{ 
$^b$) Ages for pulsars are estimated as $P/(2\dot P)$,}\\ 
\multicolumn{7}{l}{ 
      for RX J1856 the estimate of its age comes from kinematical 
      considerations.}\\ 
\multicolumn{7}{l}{ 
$^c$) Distance to PSR B1055-52 is uncertain ($\sim$ 0.9-1.5 kpc)}\\ 
\multicolumn{7}{l}{ 
$^d$) Total count rate (blackbody + non-thermal)}\\ 
\multicolumn{7}{l}{ 
$^e$) Distances determined through parallactic measurements}\\ 
\multicolumn{7}{l}{ 
$^f$) Distances determined with dispersion measure}\\ 
\multicolumn{7}{l}{ 
[1] Treves et al. (2000) ; [2] Kaplan et al. (2002); [3] Zane et al. (2002);}\\ 
\multicolumn{7}{l}{ 
[4] Hambaryan et al. (2001); [5] Haberl, Zavlin (2002); [6] Zampieri et al. 
(2001);}\\ 
\multicolumn{7}{l}{ 
[7] Becker, Trumper (1997); [8] Mirabal, Halpern (2001); [9] Pavlov et al. 
2001;}\\ 
\multicolumn{7}{l}{ 
[10] ATNF Pulsar Catalogue (see Hobbs et al. 2003); [11] Haberl et al. 2004}\\ 
\hline 
\end{tabular} } 
\vspace*{-13pt} 
\end{table} 

The main output of these calculations can be formulated as follows:
observed cooling isolated NSs originated in the Gould Belt. This result
does not depend on the uncertainties of the population synthesis model.
I want to note, that according to our calculations there should be
about several tens more unidetified 
isolated NSs with fluxes detectable by ROSAT. These objects
are hiding in crowded regions close to the galactic plane (and to the plane
of the Gould Belt). In addition there should be many NSs with ages above 
$\sim$ 1 Myr, which now are too cold now to be detected. A total expected
number of NSs depending on their ages and distances can be easily shown on the 
age-distance diagram.

If all uncertainties like initial spatial distribution of NSs,
their kick velocity distribution, characteristics of emission
(atmospheric effects etc.), mass spectrum of NSs are fixed, then
Log N - Log S distribution of close-by NSs can be a powerful tool
to put constraints on models of thermal evolution of NSs.

\subsection{Age-distance diagram}

To illustrate the population of young close-by NSs it is convenient to use
the age-distance diagram (Popov 2004).

If we are speaking about observations of thermal emission of young 
cooling NSs then most of their properties depend on their age, and
their detectability in X-rays obviously strongly depends on the distance 
(not only because of the flux dilution, but also because of the strong 
interstellar absorption of soft X-rays).
In that sense it is useful to plot an age-distance diagram (ADD) for these
objects.

There are several reasons to introduce ADDs.
First an ADD easily illustrates distributions of sources in age and 
distance. Then it is possible to plot ``expectation lines'' 
for an abundance of sources of different types and compare them with data
(i.e. line which show how many sources of smaller than a given value
we can expect to find at distances smaller than a given one).
Finally since the observability of sources depends mainly on their ages
and distances, it is possible to illustrate observational limits.

\begin{figure}[t]
\epsfig{file=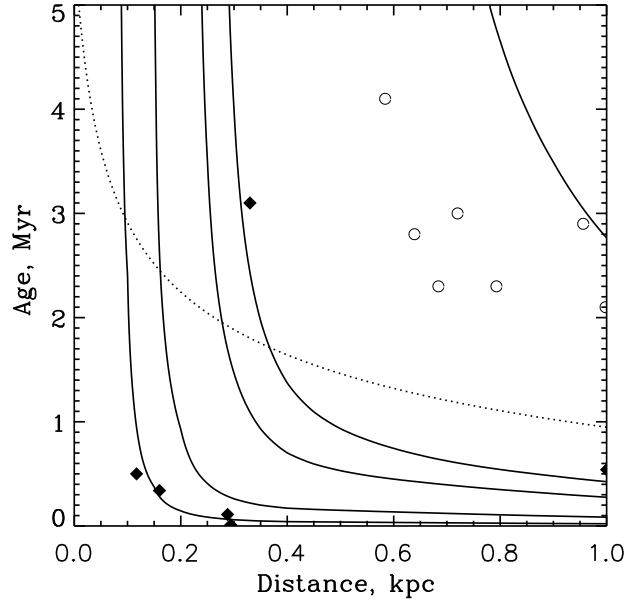, width=0.8\textwidth}
\caption{Age-distance diagram with dynamical effects.
Solid lines (from bottom to the top) corresponds to 1, 4, 13, 20 and
100 sources. The dotted line is the `visibility'' line (see details in
Popov 2004).}                          
\label{agedist}
\end{figure}

In the fig. \ref{agedist} an ADD for close-by young NSs is presented.
These objects can be observed in soft X-rays due to their thermal emission.
The thermal evolution strongly depends on the mass. 
According to different models
(see Kaminker et al. 2002, Blaschke et al. 2004
and references there in) a NS with a mass $\sim 1.3$--$1.4\, M_{\odot}$
remains hot ($T\sim (0.5$--$1)\, 10^6$~K, or equivalently $\sim$ 50--100 eV) 
up to $\sim$~0.5--1 Myr. 
It corresponds to a luminosity $\sim10^{32}$~erg~s$^{-1}$.
Usually NSs with low masses (1--1.3 $M_{\odot}$) cool down slower,
while with higher masses faster. 

To plot an ADD it is necessary to fix maximal values for age and distance
for selection of sources.
I choose 4.25 Myrs as a limiting age for selection of observed sources. 
This is the time after which even a 
low-mass NSs cools down to $\sim 10^5$~K and becomes nearly undetectable
in X-rays (see Kaminker et al. 2002, Popov et al. 2004).
A limiting distance is taken to be equal to 1 kpc (because of the absorption
it is difficult to detect in X-rays an isolated NS at larger distance
unless the star is very hot). 
There are 20 sources of different nature which are supposed to be
young close-by NSs (age $<$ 4.25 Myr, distance $<$ 1 kpc) 
(see Popov et al. 2004 and the table above): 
the M7, Geminga and the geminga-like source RX J1836.2+5925,
four PSRs with detected thermal emission (Vela, PSR 0656+14, PSR 1055-52,
PSR 1929+10\footnote{It should be mentioned, that soft X-ray emission of 
PSR 1929+10 can be due to polar caps or due to non-thermal mechanism.}), 
and seven PSRs without detection of thermal emission.
Not for all of these sources 
there are good estimates of ages and/or distances
(especially for the M7).
 In the figure there are data points for 13 objects 
for which such estimates exist (distance to PSR B1055-52 is uncertain, 
and we accept it to be 1 kpc).

Note, that there are also two PSRs with distance $<$1~kpc and with ages
in between 4.25 and 5 Myr (PSR B0823+26, PSR B0943+10) and PSR J0834-60
with distance 0.49~kpc and unknown age. These three objects
are not included into the figures. Also  PSR B1822-09
with age 0.23~Myr and distance 1~kpc is not plotted.

Two types of objects are distinguished on the graph: 
detected ones (shown as black diamonds) and undetected due to
the thermal X-ray emission (remember, there are at least 
seven additional sources -- 
six from the M7 and one geminga-like object -- 
for which there are no definite determinations of age or/and distance). 

In the  picture the ``expectation'' lines are shown 
with an inclusion of the dynamical evolution. 
I.e. here  NSs' movements in the 
Galactic potential are accurately calculated. 
To plot them we use the same model of progenitor distribution 
and SN rate as in
(Popov et al. 2003). At birth NSs obtain kick velocities in accordance
with Arzoumanian et al. (2002) model, and move in the galactic potential.

Five  ``expectation'' lines (solid) 
are plotted for 1, 4,  13, 20 and 100 sources. 
For example each point on the line  for 20 sources shows us values
of age ($t_{20}$) and distance ($R_{20}$) such that according to our model
we expect to see 20 sources with ages less than   $t_{20}$ at distances
less than $R_{20}$.

The dotted line represents the ``visibility'' line.
The idea of adding such a line
is to show the maximal distance for a given age (or vice versa
a maximal age for a given distance) at which a hot (i.e. low-mass) NS can be 
detected. To calculate this line I take the
 cooling curve  from (Kaminker et al. 2002) for 
$M=1$--$1.3\, M_{\odot}$ (in the model of these authors cooling of NSs with
$M < 1.35\, M_{\odot}$ is nearly mass-independent).
Such curves were used for example in (Popov et al. 2003). 
As soon as a cooling curve is fixed then the age determines the luminosity
of the object.
The limiting unabsorbed 
flux is assumed to be $10^{-12}$~erg~cm$^{-2}$~s$^{-1}$.
According to 
WebPIMMS\footnote{http://heasarc.gsfc.nasa.gov/docs/corp/tools.html} 
it corresponds to $\sim0.01$ 
ROSAT PSPS counts per second
for $N_H=10^{21}$~cm$^{-2}$ and a blackbody spectrum with $T=90$~eV,
or to $\sim0.1$ 
ROSAT PSPS counts per second
for $N_H=10^{20}$~cm$^{-2}$ and a blackbody spectrum with $T=50$~eV.
The latter values corresponds to the dimmest source among  the
M7 -- RX J0420.0-5022;
the former to possibly detectable hot far away objects.
Without any doubt such a simplified approach
underestimates the absorption at large distances. So
the age at which a NS is still observable is overestimated, but for
distances $<$~1~kpc and ages $<$~1~Myr 
it should not be a dramatic effect.

The fact that the line for 20 sources 
lies below  $\sim$1/2 of the observed
points tells us that according to our model not all NSs are observed.
For example, at R=1~kpc we expect to see 20 sources with ages $\sim$0.5~Myr,
but in reality we observe 
20 sources with age $<$4.25~Myr. It means that for a constant NS
formation rate we see only about 10\% of them. 
One has to bear in mind that for
seven  sources ages or/and distances are unknown, 
but they are suspected to be young and close, 
this is why here we discuss 20 and not 13 sources. 
Their inclusion or exclusion 
changes our conclusions quantitatively, but not qualitatively.
Six of the M7 sources which are not plotted on the graphs due to lack of
good distance measurement 
should populate the left bottom corner of the graphs,
since according to models of NS thermal evolution (see for example
Kaminker et al. 2002, Blaschke et al. 2004)
their ages are expected to be $<$~1~Myr, and their distances should be
$<$~500~pc.
NSs can easily escape detection as {\it coolers} simply because
they cooled down in $\sim$~1~Myr. A fraction of PSRs can be undetected
due to beaming (however, they may be observed as EGRET sources).

PSR B1929+10 lies above the "visibility line".
However it is unclear if X-ray emission of this faint 
(0.012 ROSAT cts~s$^{-1}$)
source is due to non-thermal mechanism or not (Becker \& Trumper 
1997).  Now this object is not widely accepted as a {\it 
cooler} and usually it is not plotted on the $T-t$ 
(temperature-age) graphs with cooling
curves (see for example Kaminker et al. 2002 and Blaschke et al. 2004).

Surprisingly fluctuations in the part of the plot
with small number of NSs (bottom left) are not large: four detected
{\it coolers} lie just below the line for four sources.
Probably we see nearly all low-massive NSs with ages $<$~1~Myr at
$R<$~300--400~pc.
It should be noted that we are very lucky to have such a young and close
object as the Vela pulsar. If one believes in the mass spectrum with small
fraction of NSs with $M>1.5\, M_{\odot}$ then inevitably one has to conclude
that Vela cannot be a massive NS. It can be important in selection of
cooling models (see Blaschke et al. 2004). If in a model
Vela is explained only by a cooling curve for  $M>1.5\, M_{\odot}$ then
the model may be questionable. In fact, having such a young and
 massive NS so close is very improbable.
A similar conclusion can be made for Geminga and PSR 0656+14.

There is a deficit of sources below the "visibility" line.
These sources could be already detected as dim X-ray sources by ROSAT 
but were not identified as isolated NSs.
The limit for the number of isolated NSs
from the BSC (Bright Source Catalogue) is
about 100 sources at ROSAT count rate $>$0.05 cts~s$^{-1}$ 
(see Rutledge et al. 2003). However we do not expect
to see so many young isolated NSs due to their thermal emission.
An expected number of  {\it coolers} observable by ROSAT
is about 40 objects with ages $< $~1 Myr inside
$R< 0.5$--1 kpc (see the ``visibility'' line in the
fig.\ref{agedist} in comparison with the line for 20 sources, for example).
As it was shown by Popov et al. (2003) 
most of unidentified {\it coolers} in ROSAT 
data are expected to be located at low galactic latitudes in crowded regions.

According to the classical picture all NSs were assumed to be born
as PSRs more or less similar to Crab. Depending on the initial parameters
they were assumed to be active for $\sim 10^7$~yrs.
In the last  years this picture was significantly changed. 
Without any doubts we see a deficit of young PSRs in the solar vicinity
in comparison with an expected value of young NSs from the Gould Belt and
the beaming factor cannot be the only reason for this deficit.  
This deficit  can be explained if one assumes
that a significant part of NSs do not pass through the radio  pulsar stage
or that this stage is extremely short for them (long periods of NSs from the
Magnificent seven prove it).
Of course fluctuations (in time and space) of NS production rate 
can be important as far as statistics is not that high.

I conclude this subsection with a note
 that an ADD can be the useful tool for an illustration of the
properties of close-by NSs. Its modifications can be applied to other types
of sources. For example an addition of the third axis 
(for $p$ or $\dot p$ for example) 
can be useful in discussing the population of radiopulsars.

\section{Conclusions}

 I want to stress once again that the region of $\sim$ few hundred parsecs 
around the Sun is characterized by an enhanced SN rate. This is due to the
Gould Belt. The Belt (as a structure formed by bright stars)
was well known since 19th century. In addition now we observe
young close-by NSs which, as it was shown by population synthesis calculations,
originated in the Belt. Some of EGRET unidentified sources also 
are suspected to be young NSs born in the Belt.   
 
 The existence of the Gould Belt should inevitably influence
properties of CRs detected on Earth.


%

 
\begin{acknowledgments} 
I thank the Organizers for hospitality and partial financial support.
Special thanks to many participants for stimulating discussions.
I thank Mirian Tsulaia for comments on the text.
\end{acknowledgments}


\begin{chapthebibliography}{1} 
\bibitem{}Arzoumanian, Z., Chernoff, D.F., Cordes, J.M.
2002). ``The velocity distribution of isolated radio pulsars'', ApJ 568, 289.
\bibitem{} Becker, W., Trumper, J. (1997).
``The X-ray luminosity of rotation-powered neutron stars'',
A\&A 326, 682.
\bibitem{}Berghofer, T.W., Breitschwerdt, D. (2002).
``The origin of the young stellar population in the solar neighborhood -- 
A link to the formation of the Local Bubble?'', A\&A 390, 299.
\bibitem{}Blaschke, D., Grigorian, H., Voskresensky, D. (2004).
``Cooling of neutron stars. Hadronic model.'', A\&A (in press) 
[astro-ph/0403170].
\bibitem{}Chernavskii, D.S. (2000), 
``The origin of life and thinking from the modern physics point of view'',
Physics Uspekhi 170, 157. 
\bibitem{}Dermer, C.D. (2004).
``Gamma ray bursts, sepernovae, and cosmic ray origin'',
in: Proc. of ISCRA 13th Course, Eds. M. Shapiro, T. Stanev and J. Wefel, 
World Scientific Publ., p.189.
\bibitem{}Erlykin, A.D., Wolfendeil, A.W. (1997).
``A single source of cosmic rays'', J. Phys. G 23,  979.
\bibitem{}Grenier, I.A. (2003).
``Unidentified EGRET sources in the Galaxy'',
astro-ph/0303498.			
\bibitem{}Grenier, I.A. (2004).
``The Gould Belt, star formation, and the local interstellar medium'',
astro-ph/0409096.
\bibitem{} Haberl F., Zavlin V. (2002).
``XMM-Newton observations of the isolated neutron star RX J0806.4-4123 '',
A\&A 391, 571.
\bibitem{}Haberl, F. et al. (2004).
``The isolated neutron star X-ray pulsars RX J0420.0-5022 and 
RX J0806.4-4123: New X-ray and optical observations'',
A\&A  424, 635.
\bibitem{}Hambaryan, V., Hasinger, G., Schwope, A. D., Schulz, N. S. (2001).
``Discovery of 5.16 s pulsations from the isolated neutron star RBS 1223'',
A\&A 381, 98.
\bibitem{}Hobbs, G. Manchester, R. Teoh, A. Hobbs, M. (2003).
``The ATNF Pulsar Catalogue'', 
in: Proc. of IAU Symp. 218.
"Young neutron stars and their environment"
[astro-ph/0309219].
\bibitem{}Kaminker, A.D.,  Yakovlev, D.G., Gnedin, O.Y. (2002).	
``Three types of cooling superfluid neutron stars: theory and
observations'', A\&A 383, 1076. 		
\bibitem{}Kaplan, D.L., van Kerkwijk, M.H., Anderson, J. (2002).
``The parallax and proper motion of RX J1856.5-3754 revisited'',
ApJ, 571, 447.
\bibitem{}Meszaros, P., Waxman, E. (2001).
``TeV Neutrinos from Successful and Choked Gamma-Ray Bursts'',
Phys. Rev. Lett. 87, 171102.
\bibitem{} Mirabal, N., Halpern, J.P. (2001).
``A neutron star identification for the high-energy gamma-ray source 3EG
J1835+5918 detected in the ROSAT All-Sky Survey'',
ApJ 547, L137.	
\bibitem{}Pavlov, G. et al. (2001).
``The X-ray spectrum of the Vela pulsar resolved with the Chandra X-ray
observatory'',
ApJ 552, L129.
\bibitem{}Pohl, M.,  Perrot, C.,  Grenier, I.,  Digel, S. (2003).
``The imprint of Gould's Belt on the local cosmic-ray electron spectrum'',
A\&A 409, 581.
\bibitem{}Popov, S.B., Turolla, R., Prokhorov, M.E.,  Colpi, M., Treves, A.
(2003).
``Young close-by neutron stars: the Gould Belt vs. the galactic disc'',
astro-ph/0305599.			
\bibitem{}Popov, S.B., Turolla, R., Prokhorov, M.E.,  Colpi, M., Treves, A.
(2004). ``Young compact objects in the solar vicinity'', 
in: Proc. of ISCRA 13th Course, Eds. M. Shapiro, T. Stanev and J. Wefel, 
World Scientific Publ., p.101.
\bibitem{}Popov, S.B., Turolla, R. (2004).
``Isolated neutron stars: An astrophysical perspective'', 
in: Proc. of the NATO Advanced Research Workshop On Superdense QCD Matter
and Compact Stars (in press) [astro-ph/0312369].
\bibitem{}Popov, S.B. (2004).
``Age-Distance diagram for close-by young neutron stars'', astro-ph/0407370.
\bibitem{}Prokhorov, M.E., Popov, S.B. (2002).
``Close young isolated black holes'', 
Astronomy Letters 28. 609.
\bibitem{}Rutledge, R.E., Fox, D.W., Bogosavljevic, M., Mahabal, A. (2003).
``A limit on the number of isolated neutron stars detected in the ROSAT
Bright Source Catalogue'', astro-ph/0302107.
\bibitem{}Schwope, A.D., Hasinger, G., Schwarz, R., Haberl, F., Schmidt, M.
(1999). ``The isolated neutron star candidate RBS 1223 (1RXS
J130848.6+212708)'', A\&A 341, L51.
\bibitem{}Shaviv, N. (2004). 
``Cosmic ray diffusion in the Milky Way: 
model, measurement and terrestrial effects'', 
in: Proc. of ISCRA 13th Course, Eds. M. Shapiro, T. Stanev and J. Wefel, 
World Scientific Publ., p.113.
\bibitem{} Treves, A., Turolla, R., Zane, S., Colpi, M. (2000). 
``Isolated neutron stars: accretors and coolers'',
PASP 112, 297.
\bibitem{}Tsarev, V.A. (1999). 
``Chiral effects of neutrino after supernova explosions'',
Bulletin of the Lebedev Physics Institute  2, 22.
\bibitem{}Vanbeveren, D., De Donder, E. (2003).
``The effect of the process of star formation on the temporal evolution of 
the WR and O-type star populations in the Solar neighbourhood'', 
astro-ph/0309104.
\bibitem{} Zampieri, L. et al. (2001).
``1RXS J214303.7+065419/RBS 1774: a new isolated neutron star candidate'',
A\&A 378, L5.
\bibitem{}Zane, S. et al. (2002).
``Timing analysis of the isolated neutron star RX J0720.4-3125'',
MNRAS 334, 345.
 
\end{chapthebibliography}
\end{document}